\documentclass{cjaa}      
\usepackage{graphicx,times}    
\usepackage{color}
\usepackage{colordvi}

\begin{document}
   \title{Reconstruction of Gas Temperature and Density Profiles of the Galaxy
Cluster RX J1347.5--1145}
 \volnopage{Vol.\ 8 (2008), No. 6,~ 671--676}
   \setcounter{page}{1}

   \author{Qiang Yuan\inst{1,2}
   \and Tong-Jie Zhang\inst{1,3}\mailto{}
   \and Bao-Quan Wang\inst{4}
      }
   \institute{Department of Astronomy, Beijing Normal University,
Beijing, 100875, China\\\email{tjzhang@bnu.edu.cn}
        \and
             Key Laboratory of Particle Astrophysics, Institute
of High Energy Physics, Chinese Academy of Sciences,
Beijing 100049, China\\
        \and
             Kavli Institute for Theoretical Physics China,
Institute of Theoretical Physics, Chinese Academy of Sciences
(KITPC/ITP-CAS), Beijing 100080, China\\
        \and
             Department of Physics, Dezhou University, Dezhou 253023,
China}
   \date{Received~~2008 January 7; accepted~~2008~~February 28}

\abstract{
We use observations of Sunyaev-Zel'dovich effect and X-ray 
surface brightness to reconstruct the radial profiles of gas 
temperature and density under the assumption of a spherically symmetric 
distribution of the gas. The method of reconstruction, first raised by 
Silk \& White, depend directly on the observations of the
Sunyaev-Zel'dovich effect and the X-ray surface brightness,
without involving additional assumptions such as the 
equation of state of the gas or the conditions of hydrostatic equilibrium.
We applied this method to the cluster RX J1347.5--1145, which has both the 
Sunyaev-Zel'dovich effect and X-ray observations with relative high 
precision. It is shown that it will be an effective method to 
obtain the gas distribution in galaxy clusters. Statistical errors of 
the derived temperature and density profiles of gas were estimated 
according to the observational uncertainties.
}
\authorrunning{Q. Yuan, T.-J. Zhang \& B.-Q. Wang}
\titlerunning{Temperature and Density Profiles of Galaxy
Cluster RX J1347.5-1145}
\maketitle

\section{Introduction}
Galaxy cluster is known to be the largest virial gravitational bound
system in the universe. The strong gravitational potential heats the
gas (Hydrogen and Helium) to be fully ionized with a very high
temperature $\sim10^7-10^8$K. The thermal electrons collide with
ions and emit bremsstrahlung radiation, which is mainly in X-ray
band, making galaxy clusters strong X-ray sources. The intensity of
thermal bremsstrahlung radiation is approximately proportional to
$T_e^{1/2}n_e^2$ (\cite{1999PhR...310...97B}). So by measuring the
X-ray surface brightness, we can obtain information about the gas
temperature and density distributions of the galaxy cluster. One of
the most popular models is the so-called isothermal $\beta$ model
(\cite{1976A&A....49..137C}), in which the temperature distribution
as a function of radius is assumed to be constant and the density
profile takes the form $n_e(R)=n_{e0}(1+R^2/R_c^2)^{-3\beta/2}$.
This simple model gives good approximations to many galaxy clusters.
However, more and more observations indicate that the temperature
distribution may not be a constant over the whole galaxy. Hughes et
al. (1988) introduced a revised model with a constant temperature in
a range $r<r_{\rm iso}$ and a decrease at large radius, while the
electron density has a truncation at a radius $r_{\rm lim}$. This
form provides an adequate description for some of the clusters. 
Further modification on the $\beta$ model has been proposed to deal 
with the existence of cooling flow in the center of the cluster 
(\cite{1984Natur.310..733F}).

In order to get rid of the degeneracy between $T_e$ and $n_e$, 
further assumptions like the hydrostatic equilibrium condition 
or the polytropic equation of state (EOS) of the electron gas
(\cite{1987ApJ...317..593C}; \cite{2000A&A...360L..43X}; 
\cite{2001ApJ...547...82W}), or the spectrometric analysis 
(\cite{2006ChJAA...6..181J}) are needed. However, the 
results are still model dependent. A realistic reconstruction is 
expected to come directly from the observations.

The inverse Compton scattering between the electrons in clusters
and the cosmic microwave background (CMB) photons, namely the
Sunyaev-Zel'dovich (SZ) effect (\cite{1969Ap&SS...4..301Z}; 
\cite{1970Ap&SS...7....3S}, b), provides a possible way to achieve 
this goal. The distortion of the CMB spectrum is related to the 
electron temperature and density, and will provide another equation 
of $T_e$ and $n_e$. Combining the SZ effect and the X-ray observations, 
one can obtain the electron temperature and density profiles. This 
method was first suggested by \cite{1978ApJ...226L.103S}, but has not 
yet been put into practice due to the limited instrumental sensitivity 
and resolution in the detection of the temperature fluctuation of CMB.
Up to the 1990s some preliminary observational results of the radial 
temperature distribution were obtained for a few clusters 
(\cite{1991ApJ...379..466B}; \cite{1994ApJ...420...33B}), but it was 
still difficult to obtain precise reconstructions of the temperature 
and density profiles because the data points were too sparse and uncertain
(\cite{2003ASPC..301..365W}).

In this paper, we try with this method to acquire some preliminary
information about the gas temperature and density distribution.
For simplicity, we assume a spherically symmetric distribution of  
the gas in the clusters. We adopt the $\Lambda$CDM model with 
$\Omega_M=0.27$, $\Omega_{\Lambda}=0.73$, and Hubble constant $H_0=71$ 
km s$^{-1}$ Mpc$^{-1}$ (\cite{2003ApJS..148..175S}). This paper is 
organized as follows: we describe the method of reconstruction
in Section~\ref{method}. In Section~\ref{apply} we apply this method
to the galaxy cluster RX J1347.5--1145 to reconstruct its 
temperature and density profiles, together with an estimate
of the uncertainties. The conslusions and discussions are 
presented in Section~\ref{cd}.

\section{Method of Reconstruction}
\label{method}
The fluctuation of the temperature of CMB due to the thermal SZ effect is
(\cite{1999PhR...310...97B})
\begin{equation}
\frac{\Delta T_{CMB}(R)}{T_{CMB}}=g(x)y(R),
\label{sz}
\end{equation}
where $g(x)=x\coth(x/2)-4$ with $x=h\nu/k_BT_{CMB}$ the dimensionless 
frequency, $y(R)$ is the Comptonization parameter as a function of 
the two-dimensional skymap radius $R$. In the Rayleigh-Jeans limit, 
$x\ll 1$, $g(x)=-2$, so we have $\Delta T_{CMB}(R)/T_{CMB}=-2y(R)$. 
The Comptonization parameter is related to the electron temperature 
and density by
\begin{equation}
y(R)=2A_y\int_{R}^{\infty}T_e(r)n_e(r)\frac{r{\rm
d}r}{\sqrt{r^2-R^2}},\label{y}
\end{equation}
where $A_y=k_B\sigma_T/m_ec^2$, $k_B$ is the Boltzmann constant,
$\sigma_T$ the Thomson cross section, $m_e$ the mass of electron and 
$c$ the speed of light. $r$ denotes the three-dimensional spatial 
radius from the center of the galaxy cluster, while $R=d_A\theta$ 
with $d_A$ the angular diameter distance from the Earth to the 
cluster and $\theta$ the (projected) angular separation. 
$T_e(r)$ and $n_e(r)$ represent the electron gas temperature and 
number density as functions of the radius $r$, respectively.

The X-ray surface brightness of thermal bremsstrahlung emission is
\begin{equation}
S_x(R)=\frac{1}{4\pi
(1+z)^4}\cdot2A_x\int_{R}^{\infty}T_e^{1/2}(r)n_e^2(r)C(T_e)
\frac{r{\rm d}r}{\sqrt{r^2-R^2}},
\label{xray}
\end{equation}
where $A_x=\frac{2^4e^6}{3\hbar m_ec^2}(\frac{2\pi
k_B}{3m_ec^2})^{1/2}\mu_e \bar{g}$, $\mu_e=2/(1+X)$
with $X=0.768$ the primordial Hydrogen mass fraction, $\bar{g}\approx1.2$
the average Gaunt factor (this average leads to an uncertainty of
$\sim20\%$ in the bolometric emissivity, see e.g.,
\cite{2000MNRAS.311..313E}) and $z$ the redshift of the cluster. 
$C(T_e)=\exp[-E_{\rm min}(1+z)/k_BT_e]-\exp[-E_{\rm max}(1+z)/k_BT_e]$ 
is a correction factor from the whole band of thermal bremsstrahlung 
emission to the detector bands $[E_{\rm min},E_{\rm max}]$ (e.g., 
for ROSAT, $E_{\rm min}=0.1$ keV and $E_{\rm max}=2.4$ keV).

Using Abel's integral equation, Equations (\ref{y}) and (\ref{xray}) can
be inverted to (\cite{1999ApJ...513..549Y})
\begin{eqnarray}
T_e(r)n_e(r)&=&\frac{1}{\pi A_y}\int^{r}_{\infty}\frac{{\rm
d}y(R)}{{\rm d}R}\frac{{\rm d}R}{\sqrt{R^2-r^2}},\label{tn1}\\
T_e^{1/2}(r)n_e^2(r)C(T_e)&=&\frac{4(1+z)^4}{
A_x}\int^{r}_{\infty}\frac{{\rm d}S_x(R)}{{\rm d}R}\frac{{\rm
d}R}{\sqrt{R^2-r^2}}.\label{tn2}
\end{eqnarray}
Thus, given the observational distributions of $y(R)$ and $S_x(R)$, we
can easily find the electron temperature and number density
distributions from Equations (\ref{tn1}) and (\ref{tn2}).

\section{Application to Cluster RX J1347.5--1145}
\label{apply}
RX J1347.5--1145 is a highly X-ray luminous and dynamically relaxed
galaxy cluster with redshift $z=0.451$. Both $ROSAT$ and $Chandra$ 
have measured the X-ray emission of this cluster with high precision
(\cite{1997A&A...317..646S}; Allen et al. 2002). The X-ray surface
brightness profile can be parameterized following the conventional 
$\beta$ model as
\begin{equation}
S_x(R)=S_0\left(1+\frac{R^2}{d_A^2\theta_{cx}^2}\right)
^{-3\beta_{cx}+1/2}, \label{Sx}
\end{equation}
where $d_A$ is the angular diameter distance of the cluster. For our 
adopted cosmological model, we find $d_A=1185$ Mpc. Using the $18.9$ 
ks $Chandra$ ACIS exposure, \cite{2002MNRAS.335..256A} gave a 
detailed X-ray image  of this cluster in the energy band $0.3-7.0$ 
keV. After subtracting the south-east excess, the X-ray surface 
brightness can be well described by Equation (\ref{Sx}).

Detections of the SZ effect at different frequencies in this cluster 
were published by several groups (Pointecouteau et al. 2001; 
\cite{2001PASJ...53...57K};  \cite{2002ApJ...581...53R}). An empirical 
formula similar to the $\beta$ model can also be used to parameterize 
the observational Comptonization parameter of the SZ effect $y(R)$,
\begin{equation}
y(R)=y_0\left(1+\frac{R^2}{d_A^2\theta_{cy}^2}\right)
^{-3\beta_{cy}/2+1/2}. \label{yr}
\end{equation}
We use the fitting parameters from \cite{2001ApJ...552...42P}. The 
parameters in Equations (\ref{Sx}) and (\ref{yr}) are listed in 
Table \ref{table1}.

\begin{table}[]

\centering
\begin{minipage}[]{70mm}

\caption{Parameters used in Equations (\ref{Sx}) and
(\ref{yr}).}\end{minipage} \label{table1}\vs

  \begin{tabular}{cccc}
\hline\noalign{\smallskip}
 X-ray & $S_{x0}$(ergs s$^{-1}$ cm$^{-2}$ arcmin$^{-2}$) & $\theta_{cx}$(arcsec)
 & $\beta_{cx}$\\
       & $1.34\times10^{-10}$$^a$ & $4.29\pm0.10$$^b$
 & $0.535\pm0.003$ \\
\hline\noalign{\smallskip}
 SZ & $y_{0}$ & $\theta_{cy}$(arcsec) & $\beta_{cy}$ \\
       & $4.1^{+1.7}_{-0.4}\times10^{-4}$ & $56.3^{+12.0}_{-19.1}$
 & $0.89^{+0.26}_{-0.61}$\\
\noalign{\smallskip}\hline\noalign{\smallskip}
\end{tabular}
\parbox{100mm}
{$^a$Derived from the $Chandra$ ACIS-S detector counts, see
http://heasarc.nasa.gov/Tools/w3pimms.html.\\
$^b$Note that the angular diameter distance is different from
\cite{2002MNRAS.335..256A} due to different cosmological models.
}\end{table}

Substituting $y(R)$ and $S_x(R)$ in Equations (\ref{tn1}) and (\ref{tn2}) by
Equations (\ref{Sx}) and (\ref{yr}), we can obtain
\begin{eqnarray}
T_e(r)n_e(r)&=&\frac{1}{\sqrt{\pi} A_y}\frac{y_0(3\beta_{cy}/2-1/2)}{d_A\theta_{cy}}
\frac{\Gamma(3\beta_{cy}/2)}{\Gamma(3\beta_{cy}/2+1/2)}\left(1+\frac{r^2}
{d_A^2\theta_{cy}^2}\right)^{-3\beta_{cy}/2} \nonumber\\
&=&C_y, \label{tn3} \\
T_e^{1/2}(r)n_e^2(r)C(T_e)&=&\frac{4\sqrt{\pi}(1+z)^4}{A_x}\frac{S_{x0}(3\beta_{cx}-1/2)}
{d_A\theta_{cx}}\frac{\Gamma(3\beta_{cx})}{\Gamma(3\beta_{cx}+1/2)}\left(1+\frac{r^2}
{d_A^2\theta_{cx}^2}\right)^{-3\beta_{cx}}\nonumber \\
&=&C_x.\label{tn4}
\end{eqnarray}
Then
\begin{eqnarray}
T_e(r)^{3/2}/C(T_e)&=&C_y^2/C_x, \label{tn5}\\
n_e(r)&=&C_y/T_e(r). \label{tn6}
\end{eqnarray}
From Equation (\ref{tn5}) we know that, if $\theta_{cx}=\theta_{cy}$,
$\beta_{cx}=\beta_{cy}$, the temperature should be independent of
the radius, as expected from the isothermal $\beta$ model. If
$\beta_{cy}>\beta_{cx}$, the temperature will decrease at large
radius; while for $\beta_{cy}<\beta_{cx}$, the temperature will 
increase in the outer region of the cluster, which seems to be
unreasonable.

It is easy to calculate $T_e(r)$ and $n_e(r)$ using the parameters given 
in Table \ref{table1}. To obtain the uncertainties of the derived 
temperature and density profiles, we run a Monte-Carlo (MC) sampling 
of the parameters according to their uncertainties. The parameter is thought 
to be Gaussian distributed peaked at the central value with width of 
$1\sigma$ error-bar. However, we note that the $1\sigma$ error-bars 
are not the same in the ``$+$'' and ``$-$'' directions. So the 
distribution of a parameter is the combination of two Gaussian 
functions, connected at the peak point. For example, for a parameter 
$a={a_0}^{+\sigma_1}_{-\sigma_2}$, the width of $a>a_0$ is 
$\sigma_1$ and $a<a_0$ is $\sigma_2$, with the connecting condition 
$k_1/\sigma_1=k_2/\sigma_2$. The probability distribution function of 
parameter $a$ can be written as
\begin{equation}
p(a)=\left\{
\begin{array}{ll}
\frac{k_1}{\sigma_1}\exp(-\frac{(a-a_0)^2}{2\sigma_1^2})&\ \ a>a_0,\\
\frac{k_2}{\sigma_2}\exp(-\frac{(a-a_0)^2}{2\sigma_2^2})&\ \ a<a_0.
\end{array}
\right.
\end{equation}
Here, $k_1$ and $k_2$ can be derived according to the normalization
condition $\int p(a){\rm d}a=1$. Some cut conditions are adopted in
the MC sample. Equations (\ref{tn3}) and (\ref{tn4}) require
$\beta_{cy}>1/3$ and $\beta_{cx}>1/6$; the other parameters are
required to be greater than $zero$. Even so, some of the parameter 
combinations make Equation (\ref{tn5}) unresolved. These cases were 
also discarded.

\begin{figure}[b!!]
\centering
\includegraphics[width=13cm]{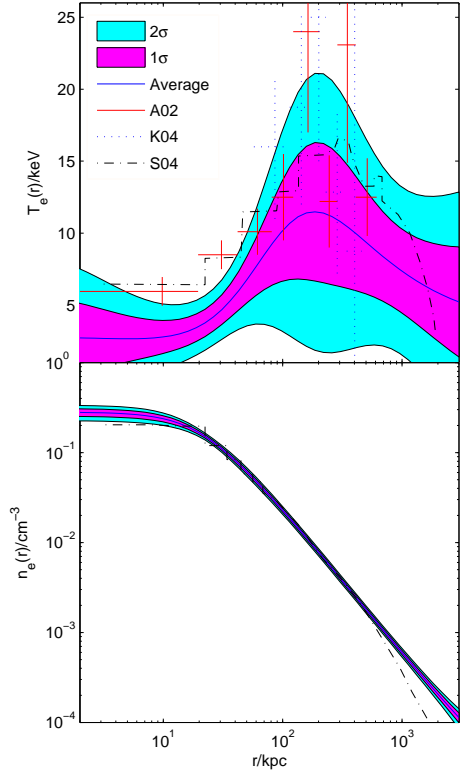}
\caption{Derived profiles of temperature (upper panel) and 
density} (lower panel) together with their uncertainties. Also shown
are the results of some previous studies, see the text for
explanations. \label{fig1eps}
\end{figure}

The average values and variances of the temperature and density of
the MC sampling are shown in Figure~\ref{fig1eps}. Shaded regions
are the $1\times$ and $2\times$ sample variances as represented by
the $\pm 1\sigma$ and $\pm 2\sigma$ errors. The results of previous
authors are also shown for comparison. The solid crosses are the
deprojected temperature profiles from the $Chandra$ X-ray
spectroscopy (Allen et al. 2002), the dotted crosses are the derived
results from the SZ effect and X-ray data using a method slightly
different from this work (\cite{2004PASJ...56...17K}), and the
dot-dashed line is the extrapolated result from the $Chandra$
spectroscopy observations (Schmidt et al. 2004). It is shown in this
figure that our derived results are roughly consistent with the
other results at $2\sigma$ level. For the inner region of the
cluster, our results show an under-estimation of the temperature.
This might be due to the systematic errors of the observations
(especially the SZ measurements). From Equations (\ref{tn5}) and
(\ref{tn6}) we know that a larger $y_0$ or a smaller $S_{x0}$ may
result in a larger central temperature. From
\cite{2001ApJ...552...42P} it was shown that different fitting
models could indeed give different $y_0$. The density profile, as
shown in the lower panel of Figure \ref{fig1eps}, is consistent with
the results in \cite{2004MNRAS.352.1413S} for $r<500$ kpc. As
pointed out in \cite{2002MNRAS.335..256A}, the surface brightness
showed a steepening with increasing radius, which means an increase
of the $\beta$ parameter. A broken power law with $\beta=0.54$
changing to $0.78$ at $r=487$ kpc could well describe the
observations. We adopt a uniform $\beta$ parameter here, so for
large radii our results are somewhat high. According to Equation
(\ref{tn5}), a larger $\beta_{cx}$ will also lead to a higher
temperature at large radius.

\section{Conclusions and Discussion}
\label{cd}
Using the observational data of the SZ effect and X-ray surface 
brightness, we applied the method suggested by 
\cite{1978ApJ...226L.103S} to the galaxy cluster RX J1347.5--1145,
to reconstruct its gas temperature and density profiles.
This is a direct way to obtain the temperature and density profiles 
of galaxy clusters, without additional theoretical assumptions. 
However, the quality of the observational data strongly affect 
the reconstruction results. Our attempt on cluster RX J1347.5--1145, 
which has both the X-ray and SZ measurements with relative high precision,
demonstrates the effectiveness of this method. The derived results 
show similar behaviors as the previous studies. It indicates 
that there is a cooling flow in the center of the cluster, but the central
value we derived is lower than the others. Poor quality and possible 
systematic uncertainties of the SZ effect data might be responsible for
this discrepancy. It should be noted that the model parameters of
the X-ray surface brightness and SZ effect were derived from finite 
area images around the center of the cluster, then extrapolated to 
large radii. This may result in additional uncertainty (see the 
discussion in Section~\ref{apply}). When high quality measurements of
the SZ effect become available in the future, it will be a 
powerful tool to study the intrinsic gas distribution independent 
of any theoretical arguments.

\begin{acknowledgements}
This work was partly supported by the National Science Foundation of
China (Grants No.10473002), the Scientific Research Foundation for
the Returned Overseas Chinese Scholars, State Education Ministry and
the Scientific Research Foundation for Undergraduate of Beijing Normal
University.
\end{acknowledgements}

\label{lastpage}

\end{document}